\documentclass[aps,prl,showpacs,amssymb,nofootinbib,10pt,superscriptaddress,twocolumn,longbibliography]{revtex4-2}
\usepackage{tikz}
\usetikzlibrary{shapes,snakes,backgrounds,fit,decorations.pathreplacing}
\usepackage{bbm,times}
\usepackage{bm}
\usepackage{amsbsy}
\usepackage{amsthm}
\usepackage{amssymb}
\usepackage{amsfonts}
\usepackage{amsmath}
\usepackage[c]{esvect} 
\usepackage{dsfont} 
\usepackage{graphicx} 
\usepackage{epsfig}
\usepackage{epstopdf}
\usepackage{dsfont}
\usepackage{multibib}
\usepackage{color}
\usepackage{enumerate}
\usepackage{multirow}

\usepackage[colorlinks]{hyperref}
\makeatletter
\newcommand\org@hypertarget{}
\let\org@hypertarget\hypertarget
\renewcommand\hypertarget[2]{%
	\Hy@raisedlink{\org@hypertarget{#1}{}}#2%
}
\makeatother
\usepackage[figure,table]{hypcap}
\usepackage{MnSymbol}
\usepackage{enumerate}
\usepackage{float}
\hypersetup{
	bookmarksnumbered,   
	pdfstartview={FitH},
	citecolor={darkblue},
	linkcolor={darkred},
	urlcolor={darkblue},
	pdfpagemode={UseOutlines}}
\definecolor{darkgreen}{RGB}{50,190,50}
\definecolor{darkblue}{RGB}{0,0,190}
\definecolor{darkred}{RGB}{238,0,0}
\usepackage{soul}
\usepackage{CJKutf8}

\newcommand*\xoverline[2][0.75]{%
    \sbox{\myboxA}{$\m@th#2$}%
    \setbox\myboxB\null
    \ht\myboxB=\ht\myboxA%
    \dp\myboxB=\dp\myboxA%
    \wd\myboxB=#1\wd\myboxA
    \sbox\myboxB{$\m@th\overline{\copy\myboxB}$}
    \setlength\mylenA{\the\wd\myboxA}
    \addtolength\mylenA{-\the\wd\myboxB}%
    \ifdim\wd\myboxB<\wd\myboxA%
       \rlap{\hskip 0.5\mylenA\usebox\myboxB}{\usebox\myboxA}%
    \else
        \hskip -0.5\mylenA\rlap{\usebox\myboxA}{\hskip 0.5\mylenA\usebox\myboxB}%
    \fi}
\makeatother

\makeatletter
\renewcommand{\p@subsection}{}
\renewcommand{\p@subsubsection}{}
\makeatother

\usepackage{braket}

\usepackage{url}

\begin{document}

\title{Driving Quantum Heat Engines Beyond Classical Limits through Multilevel Coherence}

\author{Hui Wang (\begin{CJK*}{UTF8}{gbsn}王惠\end{CJK*})\textsuperscript{*}}
\affiliation{Institute for Quantum Science and Engineering, Texas A\&M University, College Station, Texas 77843, USA}
\author{Yusef Maleki \textsuperscript{*}}
\affiliation{Institute for Quantum Science and Engineering, Texas A\&M University, College Station, Texas 77843, USA}
\author{William J. Munro}
\affiliation{Okinawa Institute of Science and Technology Graduate University, Onna-son$,$ Okinawa 904-0495$,$ Japan}
\author{Marlan O. Scully}
\affiliation{Institute for Quantum Science and Engineering, Texas A\&M University, College Station, Texas 77843, USA}
\affiliation{Baylor University, Waco, TX 76798, USA}
\affiliation{Princeton University, Princeton, New Jersey 08544, USA}

\begin{abstract}
Quantum coherence provides a controllable thermodynamic resource that can raise or lower the effective temperature of a cavity mode, enabling efficiency tuning in quantum heat engines. Here, we derive analytic expressions for the effective engine temperature, demonstrating the enhanced temperature tunability achievable via $N$-level ground-state coherence. We further unify ground- and excited-state coherence within a single analytic framework, revealing their interplay as a mechanism for thermodynamic control. Such quantum resources serve as tunable parameters that enable switching between heating, cooling, and cancellation regimes, driving the effective temperature from near-zero to divergence. Ultimately, our framework connects and generalizes previous models of quantum heat engines, and we identify rubidium atoms as a promising candidate for experimentally realizing these coherence-assisted effects.



\end{abstract}

\maketitle

\footnotetext{\textsuperscript{*} huiwangph@gmail.com}
\footnotetext{\textsuperscript{*} maleki@tamu.edu}

\emph{Introduction.--} The formulation of classical thermodynamics underscores the capacity of resources and limits the efficiency of heat engines to the Carnot bound~\cite{alicki1979quantum}, dictated by the temperatures of two thermal baths. Recent advances in quantum thermodynamics~\cite{myers2022quantum,cangemi2024quantum,quan2007quantum} extend this framework by recognizing non-classical resources such as quantum entanglement~\cite{horodecki2009quantum,guhne2009entanglement} and quantum coherence~\cite{baumgratz2014quantifying,streltsov2017colloquium,lostaglio2015description} as additional means of controlling energy flow.  The incorporation of these resources requires new principles for their thermodynamic description~\cite{vinjanampathy2016quantum}. 

\begin{figure}[h]
 \includegraphics[width=0.95\linewidth]{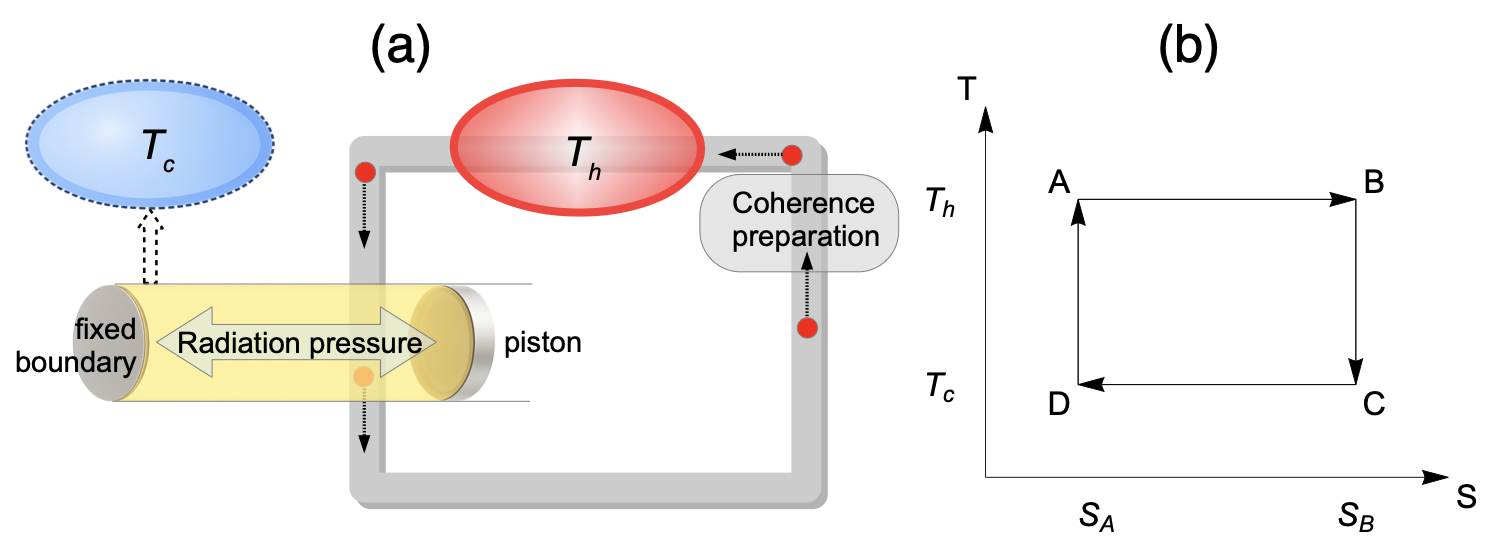} 
    \caption{(a) Quantum heat engine schematic. Radiation pressure from a thermally excited cavity field drives a piston.  Atoms that thermalize with either a hot bath at temperature $T_\mathrm{h}$ (explicitly shown) or a cold bath at temperature $T_\mathrm{c}$ (omitted for simplicity) enter the cavity, thereby controlling the effective temperature of the engine’s working medium.    (b) Carnot cycle as an idealized thermodynamic cycle performed by a heat engine, illustrated on a $T$–$S$ (temperature–entropy) diagram. The cycle takes place between a hot bath at $T_\mathrm{h}$ and a cold bath at $T_\mathrm{c}$.}
    \label{fig:qhescheme}
\end{figure}

The concept of coherence-assisted quantum heat engines (QHEs) was first introduced by Scully et al.~\cite{scully2003extracting}. In their photo-Carnot model, the piston is driven by radiation pressure, analogous to steam in classical engines, as illustrated schematically in Fig.~\ref{fig:qhescheme}(a). Thermally excited atoms, after passing through a heat bath, enter the cavity and modify the cavity field's temperature. Using incoherent two-level atoms reproduces the classical Carnot efficiency, but introducing three-level atoms with coherence between nearly degenerate ground states allows work extraction even from a single thermal reservoir, while maintaining consistency with the second law of thermodynamics~\cite{scully2003extracting,scully2011quantum,gelbwaser2015power,dorfman2013photosynthetic}. This enhancement arises from quantum interference effects~\cite{ferreri2025quantum}, which alter the balance between emission and absorption processes and effectively control the radiation temperature within the cavity. The phase and magnitude of the induced coherence thus provide powerful control parameters~\cite{scully2003extracting,amato2024heating}, enabling new mechanisms for thermodynamic optimization and engine regulation~\cite{zhang2022dynamical}.

Building on these foundations, theory and experiment have realized coherence-enabled quantum heat engines in various settings, from quantum photocells and photosynthetic complexes to superradiant platforms \cite{scully2011quantum,gelbwaser2015power,scully2010quantum,svidzinsky2011enhancing,wertnik2018optimizing,hardal2015superradiant,kim2022photonic,kim2022supercharged}. Extensions to multilevel systems and multipartite working media have clarified aspects of quantum-enhanced performance  \cite{niedenzu2015performance,turkpencce2016quantum}. Yet, key questions remain concerning the scaling of quantum enhancement with the dimensionality of the coherent manifold, and the simultaneous impact of both ground- and excited-state coherences.

In this letter, we establish a unified framework linking ground- and excited-state coherences to demonstrate that quantum coherence acts as a continuous control parameter for reversibly switching a heat engine between heating, cooling, and cancellation regimes. We also analyze how multilevel coherence, quantified by a normalized coherence parameter, governs the effective cavity temperature and efficiency. In particular, we show that coherence can either raise or lower the effective radiation temperature, and that this tunability arises from the constructive scaling of the coherence manifold.

\emph{Coherence-Assisted Engines.--} Motivated by the pioneering work~\cite{scully2003extracting}, we consider a quantum heat engine where the working medium is a single-mode radiation field confined in a cavity with perfectly reflecting mirrors. One of the mirrors behaves like a piston and moves under radiation pressure. The temperature of the cavity radiation field is regulated by hot bath atoms resonant with the field as they traverse the cavity [Fig.~\ref{fig:qhescheme} (a)]. This setup realizes a quantum photo-Carnot engine where photons serve as the working fluid.

Such an architecture can be implemented in a micromaser (or laser) system~\cite{meschede1985one}, where the cavity exhibits an exceptionally long photon lifetime, allowing even a modest flux of excited atoms to sustain quantum coherence. Also, techniques in lasing without inversion can help to generate coherence in nearly degenerate ground states, enabling stimulated emission with a small population in the excited state \cite{scully1989degenerate,kocharovskaya1992amplification}. In our setup, the engine operates within a laser cavity, where radiation pressure acts on a movable mirror functioning as a piston. The cavity field is assumed to be in thermal equilibrium with an external bath at temperature $T_{\mathrm{bath}}$, which sets the reference photon number $\bar{n}$. The pressure satisfies~\cite{scully2003extracting}
\begin{equation}
PV = \hbar \omega \bar{n},
\end{equation}
where $P$ is the radiation pressure, $V$ is the cavity volume, $\omega = m\pi c / L$ is the mode frequency for a cavity of length $L$ (with $m$ an integer), and $\bar{n}$ is the average number of thermal photons in the mode at temperature $T_{\mathrm{bath}}$.

\begin{figure}[h]
 \includegraphics[width=0.95\linewidth]{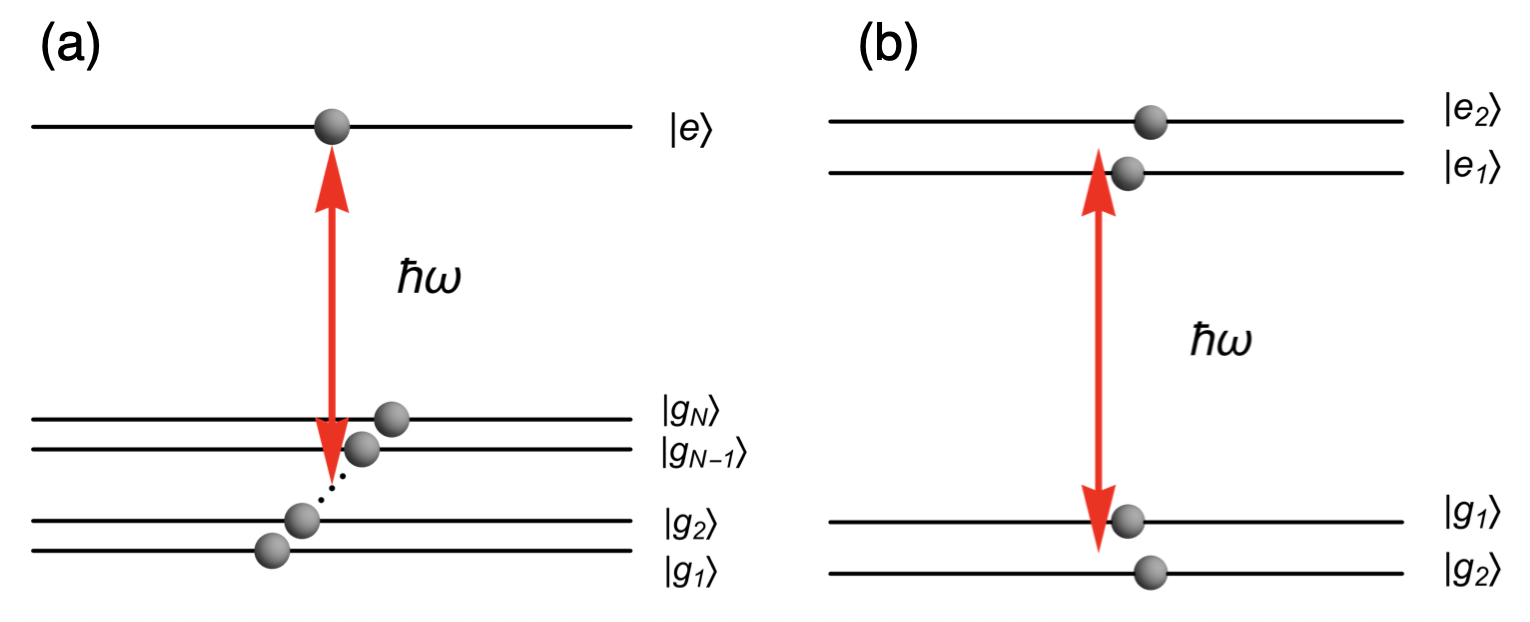} 
    \caption{Atomic configurations considered in this work. (a) Coherence among $N$ nearly degenerate ground states. (b) Four-level system with coherence between the two ground states and between the two excited states.}
    \label{fig:atomstates}
\end{figure}

We first consider a multi-level atomic system with coherence among nearly degenerate ground states. The atom possesses one excited state $|e\rangle$ and $N$ distinct ground states labeled as $|g_i\rangle$ for $i = 1, 2, \ldots, N$ . Because the ground-state splittings are small compared with the excited–ground energy gap, the manifold can be treated as effectively degenerate. Therefore, in the basis $\{|e\rangle, |g_1\rangle, |g_2\rangle, \ldots, |g_N\rangle\}$, the atom's density matrix $\rho_A$ can be expressed as 
\begin{align}
    \rho_A = P_{ee}|e\rangle\langle e|
+ \sum_{i=1}^{N}\sum_{j=1}^{N} P_{g_i g_j}\,|g_i\rangle\langle g_j| .
\end{align}
Here, $P_{ee}$ represents the population of the excited state $|e\rangle$, and $P_{g_i g_i}$ denotes the population of the ground state $|g_i\rangle$. Because the ground levels are nearly degenerate, all ground-state populations are equal, $P_{g_i g_i} = p$, where $p$ is the population probability of each degenerate ground state. Normalization then gives $P_{ee}= 1 - Np$. The off-diagonal terms $P_{g_i g_j}$ ($i\neq j$) describe coherences among the ground states, which are taken to be identical, $P_{g_i g_j}=\xi$, with $\xi$ denoting the common off-diagonal coherence amplitude. This symmetric $N$-level configuration has previously been explored but limited to the negative coherence ($\xi < 0$)~\cite{turkpencce2016quantum}. Here, we extend the analysis to include both positive and negative coherence, revealing that the sign of $\xi$ serves as a decisive switch for thermodynamic control. 

We employ the rate equation for the average cavity 
photon number $\bar{n}_\mathrm{Q}$, which is approximated by
\begin{equation}
\dot{\bar{n}}_\mathrm{Q} = \alpha \left[NP_{ee} (\bar{n}_\mathrm{Q}+1)-\sum_{i,j}P_{g_i g_{j}}\bar{n}_\mathrm{Q}\right],
\label{meaneq}
\end{equation}
where $\alpha$ denotes the rate factor (see supplementary~\cite{supplementary} for details). In the absence of atomic coherence, the steady state  average photon number follows the Boltzmann distribution and is given by $\bar{n} = (p/ { P_{ee}}-1)^{-1}$. On the other hand, with the coherence terms, we determine $\bar{n}_\mathrm{Q}$ in terms of $\bar{n}$ as
\begin{equation}
\bar{n}_\mathrm{Q} = \frac{\bar{n}} {(1+ \bar{n}\epsilon_g + \epsilon_g)} ,
\label{nqn}
\end{equation}
where we define the normalized ground-state coherence parameter $\epsilon_g$ as
\begin{equation}
\epsilon_g=\frac{\sum_{i\neq j} P_{g_i g_j}}{\sum_{i} P_{g_i g_i}}=\chi(N-1),
\label{epsilon}
\end{equation}
with $\chi=\xi/p$ defined. The allowed range for $\epsilon_g$, as derived in~\cite{supplementary}, is $-1/(\bar{n}+1)<\epsilon_g<N-1$. 


Considering the Boltzmann distribution, the effective cavity temperature set by the average photon number $\bar{n}_\mathrm{Q}$ satisfies $ k_\mathrm{B} T_\mathrm{Q} = {\hbar \omega}/{ \ln(1+\bar{n}_ \mathrm{Q}^{-1})}.$ Substituting Eq.~(\ref{nqn}) gives
\begin{equation}
    T_\mathrm{Q}= \frac{\hbar \omega }{k_\mathrm{B} \ln[(1+\epsilon_g) (1+\bar{n}^{-1})]},
    \label{tqvsepsilon}
\end{equation} 
showing that coherence effectively redefines the operational temperature of the working medium. In particular, setting $\epsilon_g=0$ in Eqs.~\eqref{nqn} and~\eqref{tqvsepsilon} recovers the classical results $\bar{n}_\mathrm{Q}=\bar{n}$ and $T_\mathrm{Q}=T_\mathrm{bath}$.

Having obtained the effective cavity temperature $T_\mathrm{Q}$ from Eq.~(\ref{tqvsepsilon}), we implement a Carnot cycle as sketched in Fig.~\ref{fig:qhescheme}(b). During the isothermal expansion ($A\!\to\!B$), the coherence-assisted atomic fuel establishes $T_\mathrm{Q}$, which can exceed $T_\mathrm{h}$ for ground-state coherence with $\epsilon_g<0$, yielding $Q_\mathrm{in}= T_\mathrm{Q}\Delta S $ and $\eta=1-T_\mathrm{c}/T_\mathrm{Q}$. Conversely, coherence with $\epsilon_g>0$ can suppress the effective temperature during isentropic compression $D\!\to\! A$, approaching $T_\mathrm{Q}\to 0$. The cycle is completed by adiabatic expansion/compression and an isothermal compression at $T_\mathrm{c}$, with $Q_\mathrm{out}=-T_\mathrm{c}\Delta S$. The efficiency takes the Carnot form $\eta=1-T_\mathrm{Q}/T_\mathrm{h}$. Thus coherence functions either to boost the hot temperature or to reduce the cold temperature, depending on its preparation. We now examine these two scenarios in detail, beginning with the case where coherence increases $T_\mathrm{h}$.

\emph{Increasing $T_\mathrm{h}$ with negative multilevel ground-state coherence.--} When hot two-level atoms heat the photon “fluid” in the cavity, the engine follows the standard Carnot efficiency $\eta=({Q_\mathrm{in}-Q_\mathrm{out}})/{Q_\mathrm{in}}=1-T_\mathrm{c}/ T_\mathrm{h} $. If the injected atoms instead possess multiple nearly degenerate ground states with negative coherence ($\epsilon_g<0$), the ground-state coherence increases the cavity photon number and raises the effective temperature to $T_\mathrm{Q}$. The resulting efficiency of this quantum engine becomes~\cite{supplementary}
\begin{equation}
    \eta_\mathrm{Q} = \eta-\frac{\ln(1+\epsilon_g)}{\ln(1+\bar{n}_\mathrm{c}^{-1})},
    \label{etaQ}
\end{equation}
where $\bar{n}_\mathrm{c}$ ($\bar{n}_\mathrm{h}$) is the average photon number of the cold (hot) bath at temperature $T_\mathrm{c}$ ($T_\mathrm{h}$). The second term in Eq.~\eqref{etaQ} represents the purely quantum coherence contribution. The expression for $\epsilon_g$ in Eq.~(\ref{epsilon}) shows that the magnitude of coherence effects increases with the number of degenerate ground states $N$, indicating that multilevel configurations amplify the influence of $\xi$ on the engine performance. For $\epsilon_g<0$ this term is positive, yielding an efficiency enhancement from coherence in the system’s internal states. In the high-temperature limit, $\bar{n}_\mathrm{Q} \approx{k_\mathrm{B} T_\mathrm{Q}}/{\hbar \omega}$ and we have $\eta_\mathrm{Q} \approx \eta-\frac{T_\mathrm{c}}{T_\mathrm{h}} \bar{n}_\mathrm{h} \epsilon_g$. In this limit, for the case $N=2$, where $P_{g_1 g_2}=|P_{g_1 g_2}|e^{i\phi}$, the efficiency simplifies to $\eta_\mathrm{Q} \approx \eta- \frac{T_\mathrm{c}}{T_\mathrm{h}} 3\bar{n}_\mathrm{h} |P_{g_1 g_2}|\cos\phi$, recovering the result obtained in~\cite{scully2003extracting}. 
Unlike earlier analyses restricted to small coherence amplitudes, here we consider the full allowed range of $\epsilon_g$, which broadens significantly as the number of coherence levels $N$ increases.

From Eq.~\eqref{etaQ} we see that by tuning the coherence parameter $\epsilon_g<0$, work can be extracted even when only a single thermal bath is present ($T_\mathrm{h}=T_\mathrm{c}$, or equivalently $\eta=0$). This striking behavior arises because quantum coherence breaks detailed balance between absorption and emission processes, biasing photon emission in favor of work extraction—an effect reminiscent of lasing without inversion~\cite{scully1989degenerate,kocharovskaya1992amplification,maruyama2009physics}. While this might appear analogous to Maxwell’s demon~\cite{leff2002maxwell,bennett1987demons,lloyd1989use}, no violation of the second law occurs, as preparing coherence constitutes an external control process whose energetic cost ensures overall thermodynamic consistency.

\begin{figure}[h]
 \includegraphics[width=\linewidth]{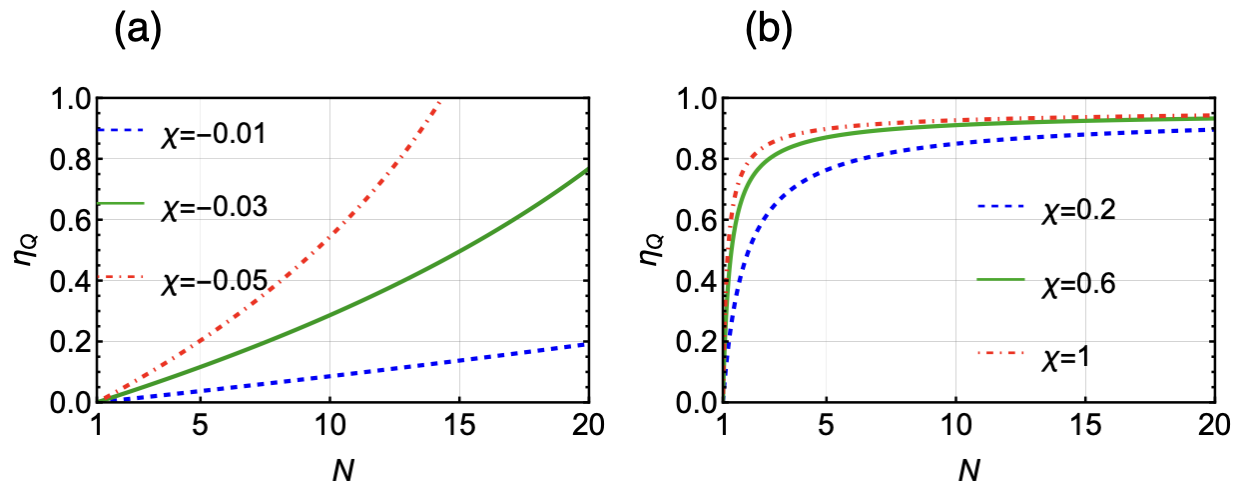} 
    \caption{ Quantum efficiency $\eta_\mathrm{Q}$ versus $N$. (a) Blue dashed, green solid and red dash-dotted  curves corresponds to $\chi = -0.01$, $\chi = -0.03$, and $\chi = -0.05$,  respectively (with  $\bar{n}_\mathrm{eq} = 0.5$). (b) Blue dashed,  green solid and red dash-dotted curves corresponds to $\chi = 0.2$, $\chi = 0.6$, and $\chi = 1$, respectively (with $\bar{n}_\mathrm{eq} = 5$). }
    \label{fig:effvsn}
\end{figure}

Considering now the extreme setting with only one thermal bath, i.e. $T_\mathrm{bath}=T_\mathrm{h}=T_\mathrm{c}$, we analyze the efficiency in the regime $\epsilon_g<0$, consistent with the above constraint. Denoting the average cavity photon number in equilibrium as $\bar{n}_\mathrm{eq}= \bar{n}_\mathrm{h}=\bar{n}_\mathrm{c}$, the quantum efficiency in Eq.~(\ref{etaQ}) becomes
\begin{equation}
    \eta_\mathrm{Q}=-\frac{\ln[1+\chi(N-1)]}{\ln(1+\bar{n}_\mathrm{eq}^{-1})},
\end{equation} 
with $\chi$ being constrained by $-1/{(N-1)(\bar{n}_\mathrm{eq}+1)} < \chi < 0$~\cite{supplementary}. Fig.~\ref{fig:effvsn}(a) shows the dependence of $\eta_\mathrm{Q}$ on both $\chi$ and $N$; here we focus on $\chi < 0$, which increases the effective temperature. We observe that as $|\chi|$ decreases, a larger value for $N$ can be explored before reaching the saturation limit $\eta_\mathrm{Q}=1$. Larger magnitudes of $\chi$ lead to a faster rise with $N$; for example, at $\chi =- 0.05$, the efficiency $\eta_\mathrm{Q}$ saturates before $N$ reaches $15$. The saturation at finite $N$ results from an overestimate of $\bar n_\mathrm{Q}$ (and consequently $T_\mathrm{Q}$) by the approximate rate equation in Eq.~(\ref{meaneq}), which is valid only in the weak-coupling regime. We note that to extend the analysis beyond this approximation, numerical analysis such as in Ref.~\cite{turkpencce2016quantum} is required.

\begin{figure}[h]
 \includegraphics[width=\linewidth]{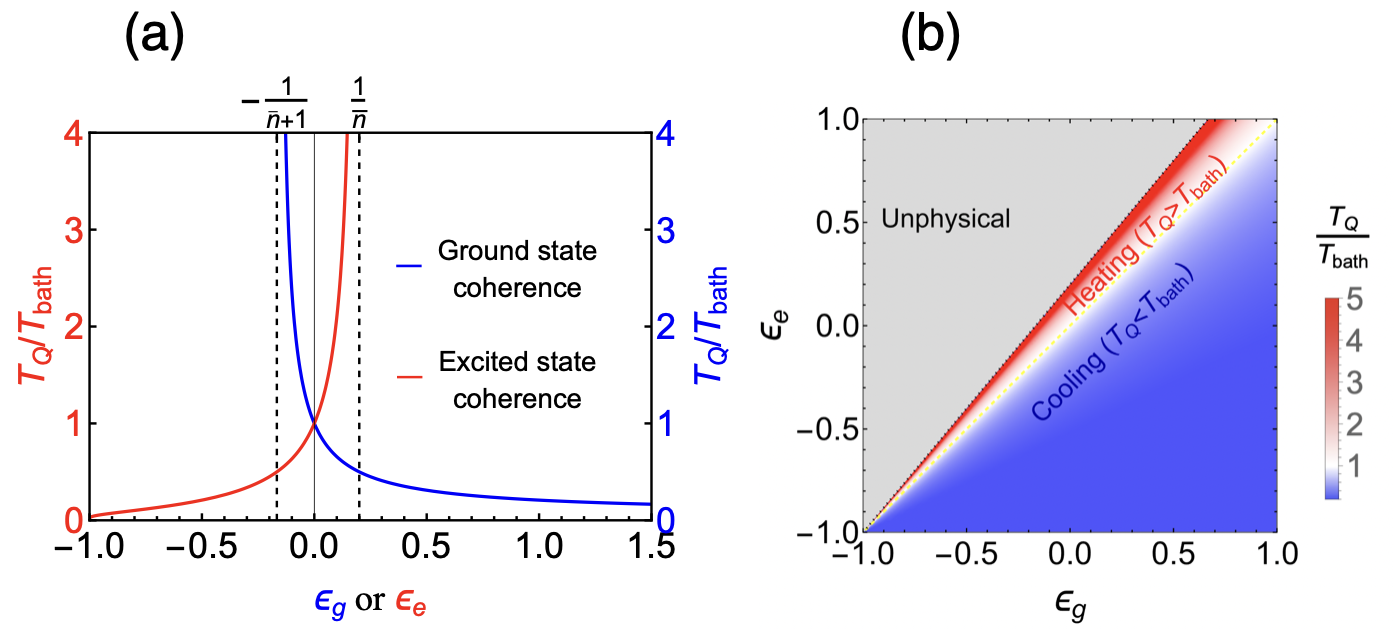} 
    \caption{ (a) Effective cavity temperature normalized to the bath temperature, $T_\mathrm{Q}/T_\mathrm{bath}$, versus ground-state ($\epsilon_g$, blue) or excited-state ($\epsilon_e$, red) coherence. Negative $\epsilon_g$ or positive $\epsilon_e$ raises the effective temperature (heating), whereas positive $\epsilon_g$ or negative $\epsilon_e$ lowers it (cooling). (b) Combined effect of ground- and excited-state coherence in the four-level configuration. Colors show $T_\mathrm{Q}/T_\mathrm{bath}$: red $>$ 1 (heating), blue $<$ 1 (cooling). In both panels, $\bar{n}=5$.}
    \label{fig:tvsepsilon}
\end{figure}

\emph{Decreasing $T_\mathrm{c}$ with positive multilevel ground-state coherence.--} So far, we have shown that $\epsilon_g < 0$ raises the cavity’s effective temperature above that of the hot bath $T_\mathrm{h}$, enabling additional work extraction. Conversely, positive coherence ($\epsilon_g > 0$) lowers the effective temperature during the isothermal compression stage of the quantum Carnot cycle. In this regime, the average photon number $\bar n_\mathrm{c}$ is replaced by the coherence-modified value $\bar n_\mathrm{Q}$ from Eq.~(\ref{nqn}), reducing the cavity temperature from $T_\mathrm{c}$ to $T_\mathrm{Q}$ and thereby decreasing the released heat $Q_\mathrm{out}$. The corresponding effective temperature, obtained from Eq.~(\ref{tqvsepsilon}), reads $ k_\mathrm{B} T_\mathrm{Q}= \hbar \omega/{\ln[(1+\epsilon_g) (1+\bar{n}_\mathrm{c}^{-1})]}.$ This mechanism effectively enhances the efficiency by cooling the working medium rather than heating it.
In principle, the term $\epsilon_g=\chi(N-1)$ can be significantly large as $N$ increases, driving the effective temperature $T_\mathrm{Q} \to 0$ [Fig.~\ref{fig:tvsepsilon}(a)] and the corresponding efficiency $\eta_\mathrm{Q} \to 1$. 

Given the physical constraint $\chi=\xi/p <1$ and the scenario of interest with positive coherence $\epsilon_g > 0$, we have $0 < \chi < 1$. Assuming a single-bath case with $T_\mathrm{bath}=T_\mathrm{h}=T_\mathrm{c}$ gives the quantum efficiency~\cite{supplementary} 
\begin{equation}
\eta_\mathrm{Q}=\left[1+\frac{\ln(1+\bar{n}_\mathrm{eq}^{-1})}{\ln[1+\chi(N-1)]}\right]^{-1},
\end{equation} 
which is a function of $\bar{n}_\mathrm{eq}$, $\chi$ and $N$. Fig.~\ref{fig:effvsn}(b) illustrates the decreasing-$T_\mathrm{c}$ mechanism: $\eta_\mathrm{Q}$ starts at zero when $N=1$ and increases steadily, approaching unity as $N$ grows to infinity. This collective enhancement arises from the enlarged ground-state manifold, with larger $\chi$ values accelerating the rise. More generally, Fig.~\ref{fig:effvsn} shows that for small $\lvert \chi \rvert$, increasing $N$ amplifies the coherence parameter $\lvert \epsilon_g \rvert$, leading to higher efficiency in both heating [$\epsilon_g<0$, panel (a)] and cooling [$\epsilon_g>0$, panel (b)] regimes.

\emph{Unified four-level configuration.--}
The analyses above considered multilevel ground-state coherence, showing that it can either raise the effective hot temperature or lower the effective cold temperature. To explore the combined effects of ground- and excited-state coherences, we introduce a four-level configuration with two nearly degenerate ground states $|g_1\rangle$, $|g_2\rangle$ and two nearly degenerate excited states $|e_1\rangle$, $|e_2\rangle$ [Fig.~\ref{fig:atomstates}(b)]. In this setup, coherence acts separately within the ground and excited manifolds while the cavity mode couples all allowed transitions. The steady-state cavity photon number is given by
\begin{equation}
\bar{n}_\mathrm{Q} = \frac{\bar{n}(1+\epsilon_e)}{1 - \bar{n}\epsilon_e + (\bar{n}+1)\epsilon_g},
\label{nqn3}
\end{equation}
where $\epsilon_e$ denotes the normalized excited-state coherence parameter, 
$\epsilon_e = (P_{e_1 e_2} + P_{e_2 e_1})/(2P_{ee})$. The corresponding effective temperature is
\begin{equation}
T_\mathrm{Q} = \frac{\hbar \omega}{k_B \ln \left( \dfrac{(\bar{n}+1)(1+\epsilon_g)}{\bar{n}(1+\epsilon_e)} \right)}.
\label{tqvsepsilon2}
\end{equation}
Details on deriving Eqs.~\eqref{nqn3} and~\eqref{tqvsepsilon2} are provided in~\cite{supplementary}.

For $\epsilon_e=0$, Eq.~(\ref{tqvsepsilon2}) reduces to the two-ground-state configuration 
[Eq.~(\ref{tqvsepsilon})], while for $\epsilon_g=0$ it reduces to the three-level configuration with two excited states and one ground state~\cite{scully2010quantum,svidzinsky2011enhancing}. In the latter case, the steady state cavity temperature is
\begin{equation}
T_\mathrm{Q} = \frac{\hbar \omega}{k_\mathrm{B} \ln\left( \frac{\bar{n} + 1}{\bar{n}(1+\epsilon_e)} \right)}.
\label{tqvsepsilon3}
\end{equation}
With $\epsilon_e$ constrained by $-1 \leq \epsilon_e < 1/\bar{n}$. 

Figure~\ref{fig:tvsepsilon}(a) compares the effective temperature ratios $T_\mathrm{Q}/T_\mathrm{bath}$ for two cases: atoms with $N$ nearly degenerate ground states and one excited state (ground-state coherence), and atoms with two nearly degenerate excited states and one ground state (excited-state coherence). Our motivation for considering only two excited states is as follows. In the $N$-ground level case, since $\epsilon_g$ is proportional to $N-1$, the temperature [Eq.~(\ref{tqvsepsilon})] can be made arbitrarily small by increasing $N$. In the two-excited-state configuration, the temperature [Eq.~(\ref{tqvsepsilon3})] approaches zero as $\epsilon_e$ decreases to its lower limit of $-1$.  Therefore, two-excited-state coherence can achieve an effective temperature of zero, analogous to the scenario of large $N$ ground-state coherence. We also remark that for negative coherence between ground states or positive coherence between excited states, $\bar{n}_\mathrm{Q}$ and $T_\mathrm{Q}$ 
increase with the magnitudes of $\epsilon_g$ and $\epsilon_e$, respectively. The divergence of $T_\mathrm{Q}$ as $\epsilon_g$ approaches its lower limit or $\epsilon_e$ approaches its upper limit reflects the breakdown of the weak-coupling approximation underlying Eq.~(\ref{meaneq}) in the high-$\bar{n}_\mathrm{Q}$ limit. However, since the qualitative trends in Fig.~\ref{fig:tvsepsilon}(a) remain physically valid, the analytical model is sufficient to capture the mechanism without requiring precise numerical evaluation.

Expression~(\ref{tqvsepsilon2}) simultaneously incorporates ground- and excited-state coherences, introducing $\epsilon_g$ and $\epsilon_e$ as independent control parameters. This grants the four-level engine a unique operational versatility: the ability to dynamically switch between heating and cooling modes by tuning the respective coherences (e.g., via engineered decoherence of either $\epsilon_e$ or $\epsilon_g$). This capability arises because the atomic state harbors competing thermodynamic drives. Fig.~\ref{fig:tvsepsilon}(b) reveals a coherence-controlled thermodynamic landscape, enabling continuous tuning between heating, cooling, and cancellation regimes. The region where $\epsilon_e > \epsilon_g$ corresponds to the heating regime (red region), while the region where $\epsilon_g > \epsilon_e$ corresponds to the cooling regime (blue region). The diagonal line $\epsilon_g = \epsilon_e$ marks the `cancellation regime' where competing resources exactly counterbalance ($T_\mathrm{Q}=T_\mathrm{bath}$). Furthermore, the figure captures the full thermodynamic spectrum: $T_\mathrm{Q}=0$ is achieved at $\epsilon_e=-1$, while $T_\mathrm{Q}$ diverges as $(\bar{n}+1)\epsilon_g-\bar n \epsilon_e \to -1$. The shadowed upper-left half-plane $1-\bar n\,\epsilon_e+(\bar n+1)\,\epsilon_g\le 0$ is unphysical, as the steady-state photon number would become negative [see Eq.~(\ref{nqn3})].


Coherence-assisted heat engines can be implemented across several experimental platforms, with rubidium (Rb) atoms offering perhaps the most immediate path to realization~\cite{oberst2007time}. In Rb, nearly degenerate Zeeman or hyperfine sublevels naturally provide controllable coherent manifolds, and coherence between these states can be precisely engineered using Raman or electromagnetically induced transparency techniques~\cite{jiang2016electromagnetically}. Strong atom–photon coupling in high-finesse optical cavities has already been demonstrated in cavity-QED systems with Rb, making it a potential candidate for implementing both four-level and $N$-level configurations~\cite{lee2014many}. NV centers in diamond, semiconductor quantum dots, and circuit-QED systems also provide complementary routes, offering tunable degeneracies, coherence control, and strong cavity coupling suitable for exploring coherence-enhanced quantum thermodynamic cycles.



In conclusion, we have shown that quantum coherence can serve as a controllable thermodynamic resource that directly modifies the effective temperature of a cavity working ``fluid''. Extending this mechanism to $N$-level systems reveals clear scaling behavior linking microscopic coherence structure to macroscopic engine performance. 
Furthermore, we established a minimal four-level configuration that unites ground- and excited-state coherence, enabling dynamic mode switching between heating, cooling, or exact cancellation within a single setup. These results demonstrate that coherence can be engineered to both enhance and suppress heat-engine efficiency, providing a quantitative route to thermodynamic control through internal quantum correlations. With strong atom–photon coupling already achieved in Rb cavity-QED and related systems, the predicted temperature-tuning effects could be observable with current experimental capabilities.

This work was funded by U.S. Department of Energy (DE-SC-0023103, DE-SC0024882); Department of Energy Contract (DE-AC36-08GO28308, SUB-2023-10388); Welch Foundation (A-1261).

\bibliography{coherenceref.bib}

\begin{thebibliography}{36}%
\makeatletter
\providecommand \@ifxundefined [1]{%
 \@ifx{#1\undefined}
}%
\providecommand \@ifnum [1]{%
 \ifnum #1\expandafter \@firstoftwo
 \else \expandafter \@secondoftwo
 \fi
}%
\providecommand \@ifx [1]{%
 \ifx #1\expandafter \@firstoftwo
 \else \expandafter \@secondoftwo
 \fi
}%
\providecommand \natexlab [1]{#1}%
\providecommand \enquote  [1]{``#1''}%
\providecommand \bibnamefont  [1]{#1}%
\providecommand \bibfnamefont [1]{#1}%
\providecommand \citenamefont [1]{#1}%
\providecommand \href@noop [0]{\@secondoftwo}%
\providecommand \href [0]{\begingroup \@sanitize@url \@href}%
\providecommand \@href[1]{\@@startlink{#1}\@@href}%
\providecommand \@@href[1]{\endgroup#1\@@endlink}%
\providecommand \@sanitize@url [0]{\catcode `\\12\catcode `\$12\catcode `\&12\catcode `\#12\catcode `\^12\catcode `\_12\catcode `\%12\relax}%
\providecommand \@@startlink[1]{}%
\providecommand \@@endlink[0]{}%
\providecommand \url  [0]{\begingroup\@sanitize@url \@url }%
\providecommand \@url [1]{\endgroup\@href {#1}{\urlprefix }}%
\providecommand \urlprefix  [0]{URL }%
\providecommand \Eprint [0]{\href }%
\providecommand \doibase [0]{https://doi.org/}%
\providecommand \selectlanguage [0]{\@gobble}%
\providecommand \bibinfo  [0]{\@secondoftwo}%
\providecommand \bibfield  [0]{\@secondoftwo}%
\providecommand \translation [1]{[#1]}%
\providecommand \BibitemOpen [0]{}%
\providecommand \bibitemStop [0]{}%
\providecommand \bibitemNoStop [0]{.\EOS\space}%
\providecommand \EOS [0]{\spacefactor3000\relax}%
\providecommand \BibitemShut  [1]{\csname bibitem#1\endcsname}%
\let\auto@bib@innerbib\@empty
\bibitem [{\citenamefont {Alicki}(1979)}]{alicki1979quantum}%
  \BibitemOpen
  \bibfield  {author} {\bibinfo {author} {\bibfnamefont {R.}~\bibnamefont {Alicki}},\ }\bibfield  {title} {\bibinfo {title} {The quantum open system as a model of the heat engine},\ }\href@noop {} {\bibfield  {journal} {\bibinfo  {journal} {Journal of Physics A: Mathematical and General}\ }\textbf {\bibinfo {volume} {12}},\ \bibinfo {pages} {L103} (\bibinfo {year} {1979})}\BibitemShut {NoStop}%
\bibitem [{\citenamefont {Myers}\ \emph {et~al.}(2022)\citenamefont {Myers}, \citenamefont {Abah},\ and\ \citenamefont {Deffner}}]{myers2022quantum}%
  \BibitemOpen
  \bibfield  {author} {\bibinfo {author} {\bibfnamefont {N.~M.}\ \bibnamefont {Myers}}, \bibinfo {author} {\bibfnamefont {O.}~\bibnamefont {Abah}},\ and\ \bibinfo {author} {\bibfnamefont {S.}~\bibnamefont {Deffner}},\ }\bibfield  {title} {\bibinfo {title} {Quantum thermodynamic devices: From theoretical proposals to experimental reality},\ }\href@noop {} {\bibfield  {journal} {\bibinfo  {journal} {AVS quantum science}\ }\textbf {\bibinfo {volume} {4}} (\bibinfo {year} {2022})}\BibitemShut {NoStop}%
\bibitem [{\citenamefont {Cangemi}\ \emph {et~al.}(2024)\citenamefont {Cangemi}, \citenamefont {Bhadra},\ and\ \citenamefont {Levy}}]{cangemi2024quantum}%
  \BibitemOpen
  \bibfield  {author} {\bibinfo {author} {\bibfnamefont {L.~M.}\ \bibnamefont {Cangemi}}, \bibinfo {author} {\bibfnamefont {C.}~\bibnamefont {Bhadra}},\ and\ \bibinfo {author} {\bibfnamefont {A.}~\bibnamefont {Levy}},\ }\bibfield  {title} {\bibinfo {title} {Quantum engines and refrigerators},\ }\href@noop {} {\bibfield  {journal} {\bibinfo  {journal} {Physics Reports}\ }\textbf {\bibinfo {volume} {1087}},\ \bibinfo {pages} {1} (\bibinfo {year} {2024})}\BibitemShut {NoStop}%
\bibitem [{\citenamefont {Quan}\ \emph {et~al.}(2007)\citenamefont {Quan}, \citenamefont {Liu}, \citenamefont {Sun},\ and\ \citenamefont {Nori}}]{quan2007quantum}%
  \BibitemOpen
  \bibfield  {author} {\bibinfo {author} {\bibfnamefont {H.~T.}\ \bibnamefont {Quan}}, \bibinfo {author} {\bibfnamefont {Y.-x.}\ \bibnamefont {Liu}}, \bibinfo {author} {\bibfnamefont {C.~P.}\ \bibnamefont {Sun}},\ and\ \bibinfo {author} {\bibfnamefont {F.}~\bibnamefont {Nori}},\ }\bibfield  {title} {\bibinfo {title} {Quantum thermodynamic cycles and quantum heat engines},\ }\href@noop {} {\bibfield  {journal} {\bibinfo  {journal} {Phys. Rev. E}\ }\textbf {\bibinfo {volume} {76}},\ \bibinfo {pages} {031105} (\bibinfo {year} {2007})}\BibitemShut {NoStop}%
\bibitem [{\citenamefont {Horodecki}\ \emph {et~al.}(2009)\citenamefont {Horodecki}, \citenamefont {Horodecki}, \citenamefont {Horodecki},\ and\ \citenamefont {Horodecki}}]{horodecki2009quantum}%
  \BibitemOpen
  \bibfield  {author} {\bibinfo {author} {\bibfnamefont {R.}~\bibnamefont {Horodecki}}, \bibinfo {author} {\bibfnamefont {P.}~\bibnamefont {Horodecki}}, \bibinfo {author} {\bibfnamefont {M.}~\bibnamefont {Horodecki}},\ and\ \bibinfo {author} {\bibfnamefont {K.}~\bibnamefont {Horodecki}},\ }\bibfield  {title} {\bibinfo {title} {Quantum entanglement},\ }\href@noop {} {\bibfield  {journal} {\bibinfo  {journal} {Reviews of modern physics}\ }\textbf {\bibinfo {volume} {81}},\ \bibinfo {pages} {865} (\bibinfo {year} {2009})}\BibitemShut {NoStop}%
\bibitem [{\citenamefont {G{\"u}hne}\ and\ \citenamefont {T{\'o}th}(2009)}]{guhne2009entanglement}%
  \BibitemOpen
  \bibfield  {author} {\bibinfo {author} {\bibfnamefont {O.}~\bibnamefont {G{\"u}hne}}\ and\ \bibinfo {author} {\bibfnamefont {G.}~\bibnamefont {T{\'o}th}},\ }\bibfield  {title} {\bibinfo {title} {Entanglement detection},\ }\href@noop {} {\bibfield  {journal} {\bibinfo  {journal} {Physics Reports}\ }\textbf {\bibinfo {volume} {474}},\ \bibinfo {pages} {1} (\bibinfo {year} {2009})}\BibitemShut {NoStop}%
\bibitem [{\citenamefont {Baumgratz}\ \emph {et~al.}(2014)\citenamefont {Baumgratz}, \citenamefont {Cramer},\ and\ \citenamefont {Plenio}}]{baumgratz2014quantifying}%
  \BibitemOpen
  \bibfield  {author} {\bibinfo {author} {\bibfnamefont {T.}~\bibnamefont {Baumgratz}}, \bibinfo {author} {\bibfnamefont {M.}~\bibnamefont {Cramer}},\ and\ \bibinfo {author} {\bibfnamefont {M.~B.}\ \bibnamefont {Plenio}},\ }\bibfield  {title} {\bibinfo {title} {Quantifying coherence},\ }\href@noop {} {\bibfield  {journal} {\bibinfo  {journal} {Physical review letters}\ }\textbf {\bibinfo {volume} {113}},\ \bibinfo {pages} {140401} (\bibinfo {year} {2014})}\BibitemShut {NoStop}%
\bibitem [{\citenamefont {Streltsov}\ \emph {et~al.}(2017)\citenamefont {Streltsov}, \citenamefont {Adesso},\ and\ \citenamefont {Plenio}}]{streltsov2017colloquium}%
  \BibitemOpen
  \bibfield  {author} {\bibinfo {author} {\bibfnamefont {A.}~\bibnamefont {Streltsov}}, \bibinfo {author} {\bibfnamefont {G.}~\bibnamefont {Adesso}},\ and\ \bibinfo {author} {\bibfnamefont {M.~B.}\ \bibnamefont {Plenio}},\ }\bibfield  {title} {\bibinfo {title} {Colloquium: Quantum coherence as a resource},\ }\href@noop {} {\bibfield  {journal} {\bibinfo  {journal} {Reviews of Modern Physics}\ }\textbf {\bibinfo {volume} {89}},\ \bibinfo {pages} {041003} (\bibinfo {year} {2017})}\BibitemShut {NoStop}%
\bibitem [{\citenamefont {Lostaglio}\ \emph {et~al.}(2015)\citenamefont {Lostaglio}, \citenamefont {Jennings},\ and\ \citenamefont {Rudolph}}]{lostaglio2015description}%
  \BibitemOpen
  \bibfield  {author} {\bibinfo {author} {\bibfnamefont {M.}~\bibnamefont {Lostaglio}}, \bibinfo {author} {\bibfnamefont {D.}~\bibnamefont {Jennings}},\ and\ \bibinfo {author} {\bibfnamefont {T.}~\bibnamefont {Rudolph}},\ }\bibfield  {title} {\bibinfo {title} {Description of quantum coherence in thermodynamic processes requires constraints beyond free energy},\ }\href@noop {} {\bibfield  {journal} {\bibinfo  {journal} {Nature communications}\ }\textbf {\bibinfo {volume} {6}},\ \bibinfo {pages} {6383} (\bibinfo {year} {2015})}\BibitemShut {NoStop}%
\bibitem [{\citenamefont {Vinjanampathy}\ and\ \citenamefont {Anders}(2016)}]{vinjanampathy2016quantum}%
  \BibitemOpen
  \bibfield  {author} {\bibinfo {author} {\bibfnamefont {S.}~\bibnamefont {Vinjanampathy}}\ and\ \bibinfo {author} {\bibfnamefont {J.}~\bibnamefont {Anders}},\ }\bibfield  {title} {\bibinfo {title} {Quantum thermodynamics},\ }\href@noop {} {\bibfield  {journal} {\bibinfo  {journal} {Contemporary Physics}\ }\textbf {\bibinfo {volume} {57}},\ \bibinfo {pages} {545} (\bibinfo {year} {2016})}\BibitemShut {NoStop}%
\bibitem [{\citenamefont {Scully}\ \emph {et~al.}(2003)\citenamefont {Scully}, \citenamefont {Zubairy}, \citenamefont {Agarwal},\ and\ \citenamefont {Walther}}]{scully2003extracting}%
  \BibitemOpen
  \bibfield  {author} {\bibinfo {author} {\bibfnamefont {M.~O.}\ \bibnamefont {Scully}}, \bibinfo {author} {\bibfnamefont {M.~S.}\ \bibnamefont {Zubairy}}, \bibinfo {author} {\bibfnamefont {G.~S.}\ \bibnamefont {Agarwal}},\ and\ \bibinfo {author} {\bibfnamefont {H.}~\bibnamefont {Walther}},\ }\bibfield  {title} {\bibinfo {title} {Extracting work from a single heat bath via vanishing quantum coherence},\ }\href@noop {} {\bibfield  {journal} {\bibinfo  {journal} {Science}\ }\textbf {\bibinfo {volume} {299}},\ \bibinfo {pages} {862} (\bibinfo {year} {2003})}\BibitemShut {NoStop}%
\bibitem [{\citenamefont {Scully}\ \emph {et~al.}(2011)\citenamefont {Scully}, \citenamefont {Chapin}, \citenamefont {Dorfman}, \citenamefont {Kim},\ and\ \citenamefont {Svidzinsky}}]{scully2011quantum}%
  \BibitemOpen
  \bibfield  {author} {\bibinfo {author} {\bibfnamefont {M.~O.}\ \bibnamefont {Scully}}, \bibinfo {author} {\bibfnamefont {K.~R.}\ \bibnamefont {Chapin}}, \bibinfo {author} {\bibfnamefont {K.~E.}\ \bibnamefont {Dorfman}}, \bibinfo {author} {\bibfnamefont {M.~B.}\ \bibnamefont {Kim}},\ and\ \bibinfo {author} {\bibfnamefont {A.}~\bibnamefont {Svidzinsky}},\ }\bibfield  {title} {\bibinfo {title} {Quantum heat engine power can be increased by noise-induced coherence},\ }\href@noop {} {\bibfield  {journal} {\bibinfo  {journal} {Proceedings of the National Academy of Sciences}\ }\textbf {\bibinfo {volume} {108}},\ \bibinfo {pages} {15097} (\bibinfo {year} {2011})}\BibitemShut {NoStop}%
\bibitem [{\citenamefont {Gelbwaser-Klimovsky}\ \emph {et~al.}(2015)\citenamefont {Gelbwaser-Klimovsky}, \citenamefont {Niedenzu}, \citenamefont {Brumer},\ and\ \citenamefont {Kurizki}}]{gelbwaser2015power}%
  \BibitemOpen
  \bibfield  {author} {\bibinfo {author} {\bibfnamefont {D.}~\bibnamefont {Gelbwaser-Klimovsky}}, \bibinfo {author} {\bibfnamefont {W.}~\bibnamefont {Niedenzu}}, \bibinfo {author} {\bibfnamefont {P.}~\bibnamefont {Brumer}},\ and\ \bibinfo {author} {\bibfnamefont {G.}~\bibnamefont {Kurizki}},\ }\bibfield  {title} {\bibinfo {title} {Power enhancement of heat engines via correlated thermalization in a three-level “working fluid”},\ }\href@noop {} {\bibfield  {journal} {\bibinfo  {journal} {Scientific reports}\ }\textbf {\bibinfo {volume} {5}},\ \bibinfo {pages} {14413} (\bibinfo {year} {2015})}\BibitemShut {NoStop}%
\bibitem [{\citenamefont {Dorfman}\ \emph {et~al.}(2013)\citenamefont {Dorfman}, \citenamefont {Voronine}, \citenamefont {Mukamel},\ and\ \citenamefont {Scully}}]{dorfman2013photosynthetic}%
  \BibitemOpen
  \bibfield  {author} {\bibinfo {author} {\bibfnamefont {K.~E.}\ \bibnamefont {Dorfman}}, \bibinfo {author} {\bibfnamefont {D.~V.}\ \bibnamefont {Voronine}}, \bibinfo {author} {\bibfnamefont {S.}~\bibnamefont {Mukamel}},\ and\ \bibinfo {author} {\bibfnamefont {M.~O.}\ \bibnamefont {Scully}},\ }\bibfield  {title} {\bibinfo {title} {Photosynthetic reaction center as a quantum heat engine},\ }\href@noop {} {\bibfield  {journal} {\bibinfo  {journal} {Proceedings of the National Academy of Sciences}\ }\textbf {\bibinfo {volume} {110}},\ \bibinfo {pages} {2746} (\bibinfo {year} {2013})}\BibitemShut {NoStop}%
\bibitem [{\citenamefont {Ferreri}\ \emph {et~al.}(2025)\citenamefont {Ferreri}, \citenamefont {Wang}, \citenamefont {Nori}, \citenamefont {Wilhelm},\ and\ \citenamefont {Bruschi}}]{ferreri2025quantum}%
  \BibitemOpen
  \bibfield  {author} {\bibinfo {author} {\bibfnamefont {A.}~\bibnamefont {Ferreri}}, \bibinfo {author} {\bibfnamefont {H.}~\bibnamefont {Wang}}, \bibinfo {author} {\bibfnamefont {F.}~\bibnamefont {Nori}}, \bibinfo {author} {\bibfnamefont {F.~K.}\ \bibnamefont {Wilhelm}},\ and\ \bibinfo {author} {\bibfnamefont {D.~E.}\ \bibnamefont {Bruschi}},\ }\bibfield  {title} {\bibinfo {title} {Quantum heat engine based on quantum interferometry: The su(1,1) otto cycle},\ }\href@noop {} {\bibfield  {journal} {\bibinfo  {journal} {Phys. Rev. Research}\ }\textbf {\bibinfo {volume} {7}},\ \bibinfo {pages} {013284} (\bibinfo {year} {2025})}\BibitemShut {NoStop}%
\bibitem [{\citenamefont {Amato}\ \emph {et~al.}(2024)\citenamefont {Amato}, \citenamefont {Pellitteri}, \citenamefont {Palma}, \citenamefont {Lorenzo},\ and\ \citenamefont {Lo~Franco}}]{amato2024heating}%
  \BibitemOpen
  \bibfield  {author} {\bibinfo {author} {\bibfnamefont {F.}~\bibnamefont {Amato}}, \bibinfo {author} {\bibfnamefont {C.}~\bibnamefont {Pellitteri}}, \bibinfo {author} {\bibfnamefont {G.~M.}\ \bibnamefont {Palma}}, \bibinfo {author} {\bibfnamefont {S.}~\bibnamefont {Lorenzo}},\ and\ \bibinfo {author} {\bibfnamefont {R.}~\bibnamefont {Lo~Franco}},\ }\bibfield  {title} {\bibinfo {title} {Heating and cooling processes via phaseonium-driven dynamics of cascade systems},\ }\href@noop {} {\bibfield  {journal} {\bibinfo  {journal} {Physical Review A}\ }\textbf {\bibinfo {volume} {109}},\ \bibinfo {pages} {043705} (\bibinfo {year} {2024})}\BibitemShut {NoStop}%
\bibitem [{\citenamefont {Zhang}\ \emph {et~al.}(2022)\citenamefont {Zhang}, \citenamefont {Zhang}, \citenamefont {Ding}, \citenamefont {Li}, \citenamefont {Bu}, \citenamefont {Wang}, \citenamefont {Yan}, \citenamefont {Su}, \citenamefont {Chen}, \citenamefont {Nori}, \citenamefont {\"Ozdemir}, \citenamefont {Zhou}, \citenamefont {Jing},\ and\ \citenamefont {Feng}}]{zhang2022dynamical}%
  \BibitemOpen
  \bibfield  {author} {\bibinfo {author} {\bibfnamefont {J.~W.}\ \bibnamefont {Zhang}}, \bibinfo {author} {\bibfnamefont {J.~Q.}\ \bibnamefont {Zhang}}, \bibinfo {author} {\bibfnamefont {G.~Y.}\ \bibnamefont {Ding}}, \bibinfo {author} {\bibfnamefont {J.~C.}\ \bibnamefont {Li}}, \bibinfo {author} {\bibfnamefont {J.~T.}\ \bibnamefont {Bu}}, \bibinfo {author} {\bibfnamefont {B.}~\bibnamefont {Wang}}, \bibinfo {author} {\bibfnamefont {L.~L.}\ \bibnamefont {Yan}}, \bibinfo {author} {\bibfnamefont {S.~L.}\ \bibnamefont {Su}}, \bibinfo {author} {\bibfnamefont {L.}~\bibnamefont {Chen}}, \bibinfo {author} {\bibfnamefont {F.}~\bibnamefont {Nori}}, \bibinfo {author} {\bibfnamefont {c.~K.}\ \bibnamefont {\"Ozdemir}}, \bibinfo {author} {\bibfnamefont {F.}~\bibnamefont {Zhou}}, \bibinfo {author} {\bibfnamefont {H.}~\bibnamefont {Jing}},\ and\ \bibinfo {author} {\bibfnamefont {M.}~\bibnamefont {Feng}},\ }\bibfield  {title} {\bibinfo {title} {Dynamical control of quantum heat engines using exceptional points},\ }\href@noop
  {} {\bibfield  {journal} {\bibinfo  {journal} {Nat. Commun.}\ }\textbf {\bibinfo {volume} {13}},\ \bibinfo {pages} {6225} (\bibinfo {year} {2022})}\BibitemShut {NoStop}%
\bibitem [{\citenamefont {Scully}(2010)}]{scully2010quantum}%
  \BibitemOpen
  \bibfield  {author} {\bibinfo {author} {\bibfnamefont {M.~O.}\ \bibnamefont {Scully}},\ }\bibfield  {title} {\bibinfo {title} {Quantum photocell: Using quantum coherence to reduce radiative recombination and increase efficiency},\ }\href@noop {} {\bibfield  {journal} {\bibinfo  {journal} {Physical review letters}\ }\textbf {\bibinfo {volume} {104}},\ \bibinfo {pages} {207701} (\bibinfo {year} {2010})}\BibitemShut {NoStop}%
\bibitem [{\citenamefont {Svidzinsky}\ \emph {et~al.}(2011)\citenamefont {Svidzinsky}, \citenamefont {Dorfman},\ and\ \citenamefont {Scully}}]{svidzinsky2011enhancing}%
  \BibitemOpen
  \bibfield  {author} {\bibinfo {author} {\bibfnamefont {A.~A.}\ \bibnamefont {Svidzinsky}}, \bibinfo {author} {\bibfnamefont {K.~E.}\ \bibnamefont {Dorfman}},\ and\ \bibinfo {author} {\bibfnamefont {M.~O.}\ \bibnamefont {Scully}},\ }\bibfield  {title} {\bibinfo {title} {Enhancing photovoltaic power by fano-induced coherence},\ }\href@noop {} {\bibfield  {journal} {\bibinfo  {journal} {Physical Review A}\ }\textbf {\bibinfo {volume} {84}},\ \bibinfo {pages} {053818} (\bibinfo {year} {2011})}\BibitemShut {NoStop}%
\bibitem [{\citenamefont {Wertnik}\ \emph {et~al.}(2018)\citenamefont {Wertnik}, \citenamefont {Chin}, \citenamefont {Nori},\ and\ \citenamefont {Lambert}}]{wertnik2018optimizing}%
  \BibitemOpen
  \bibfield  {author} {\bibinfo {author} {\bibfnamefont {M.}~\bibnamefont {Wertnik}}, \bibinfo {author} {\bibfnamefont {A.}~\bibnamefont {Chin}}, \bibinfo {author} {\bibfnamefont {F.}~\bibnamefont {Nori}},\ and\ \bibinfo {author} {\bibfnamefont {N.}~\bibnamefont {Lambert}},\ }\bibfield  {title} {\bibinfo {title} {Optimizing co-operative multi-environment dynamics in a dark-state-enhanced photosynthetic heat engine},\ }\href@noop {} {\bibfield  {journal} {\bibinfo  {journal} {J. Chem. Phys.}\ }\textbf {\bibinfo {volume} {149}},\ \bibinfo {pages} {084112} (\bibinfo {year} {2018})}\BibitemShut {NoStop}%
\bibitem [{\citenamefont {Hardal}\ and\ \citenamefont {M{\"u}stecapl{\i}o{\u{g}}lu}(2015)}]{hardal2015superradiant}%
  \BibitemOpen
  \bibfield  {author} {\bibinfo {author} {\bibfnamefont {A.~{\"U}.}\ \bibnamefont {Hardal}}\ and\ \bibinfo {author} {\bibfnamefont {{\"O}.~E.}\ \bibnamefont {M{\"u}stecapl{\i}o{\u{g}}lu}},\ }\bibfield  {title} {\bibinfo {title} {Superradiant quantum heat engine},\ }\href@noop {} {\bibfield  {journal} {\bibinfo  {journal} {Scientific reports}\ }\textbf {\bibinfo {volume} {5}},\ \bibinfo {pages} {12953} (\bibinfo {year} {2015})}\BibitemShut {NoStop}%
\bibitem [{\citenamefont {Kim}\ \emph {et~al.}(2022{\natexlab{a}})\citenamefont {Kim}, \citenamefont {Oh}, \citenamefont {Yang}, \citenamefont {Kim}, \citenamefont {Lee},\ and\ \citenamefont {An}}]{kim2022photonic}%
  \BibitemOpen
  \bibfield  {author} {\bibinfo {author} {\bibfnamefont {J.}~\bibnamefont {Kim}}, \bibinfo {author} {\bibfnamefont {S.-h.}\ \bibnamefont {Oh}}, \bibinfo {author} {\bibfnamefont {D.}~\bibnamefont {Yang}}, \bibinfo {author} {\bibfnamefont {J.}~\bibnamefont {Kim}}, \bibinfo {author} {\bibfnamefont {M.}~\bibnamefont {Lee}},\ and\ \bibinfo {author} {\bibfnamefont {K.}~\bibnamefont {An}},\ }\bibfield  {title} {\bibinfo {title} {A photonic quantum engine driven by superradiance},\ }\href@noop {} {\bibfield  {journal} {\bibinfo  {journal} {Nature Photonics}\ }\textbf {\bibinfo {volume} {16}},\ \bibinfo {pages} {707} (\bibinfo {year} {2022}{\natexlab{a}})}\BibitemShut {NoStop}%
\bibitem [{\citenamefont {Kim}\ \emph {et~al.}(2022{\natexlab{b}})\citenamefont {Kim}, \citenamefont {Scully},\ and\ \citenamefont {Svidzinsky}}]{kim2022supercharged}%
  \BibitemOpen
  \bibfield  {author} {\bibinfo {author} {\bibfnamefont {M.}~\bibnamefont {Kim}}, \bibinfo {author} {\bibfnamefont {M.}~\bibnamefont {Scully}},\ and\ \bibinfo {author} {\bibfnamefont {A.}~\bibnamefont {Svidzinsky}},\ }\bibfield  {title} {\bibinfo {title} {A supercharged photonic quantum heat engine},\ }\href@noop {} {\bibfield  {journal} {\bibinfo  {journal} {Nature Photonics}\ }\textbf {\bibinfo {volume} {16}},\ \bibinfo {pages} {669} (\bibinfo {year} {2022}{\natexlab{b}})}\BibitemShut {NoStop}%
\bibitem [{\citenamefont {Niedenzu}\ \emph {et~al.}(2015)\citenamefont {Niedenzu}, \citenamefont {Gelbwaser-Klimovsky},\ and\ \citenamefont {Kurizki}}]{niedenzu2015performance}%
  \BibitemOpen
  \bibfield  {author} {\bibinfo {author} {\bibfnamefont {W.}~\bibnamefont {Niedenzu}}, \bibinfo {author} {\bibfnamefont {D.}~\bibnamefont {Gelbwaser-Klimovsky}},\ and\ \bibinfo {author} {\bibfnamefont {G.}~\bibnamefont {Kurizki}},\ }\bibfield  {title} {\bibinfo {title} {Performance limits of multilevel and multipartite quantum heat machines},\ }\href@noop {} {\bibfield  {journal} {\bibinfo  {journal} {Physical Review E}\ }\textbf {\bibinfo {volume} {92}},\ \bibinfo {pages} {042123} (\bibinfo {year} {2015})}\BibitemShut {NoStop}%
\bibitem [{\citenamefont {T{\"u}rkpen{\c{c}}e}\ and\ \citenamefont {M{\"u}stecapl{\i}o{\u{g}}lu}(2016)}]{turkpencce2016quantum}%
  \BibitemOpen
  \bibfield  {author} {\bibinfo {author} {\bibfnamefont {D.}~\bibnamefont {T{\"u}rkpen{\c{c}}e}}\ and\ \bibinfo {author} {\bibfnamefont {{\"O}.~E.}\ \bibnamefont {M{\"u}stecapl{\i}o{\u{g}}lu}},\ }\bibfield  {title} {\bibinfo {title} {Quantum fuel with multilevel atomic coherence for ultrahigh specific work in a photonic carnot engine},\ }\href@noop {} {\bibfield  {journal} {\bibinfo  {journal} {Physical Review E}\ }\textbf {\bibinfo {volume} {93}},\ \bibinfo {pages} {012145} (\bibinfo {year} {2016})}\BibitemShut {NoStop}%
\bibitem [{\citenamefont {Meschede}\ \emph {et~al.}(1985)\citenamefont {Meschede}, \citenamefont {Walther},\ and\ \citenamefont {M{\"u}ller}}]{meschede1985one}%
  \BibitemOpen
  \bibfield  {author} {\bibinfo {author} {\bibfnamefont {D.}~\bibnamefont {Meschede}}, \bibinfo {author} {\bibfnamefont {H.}~\bibnamefont {Walther}},\ and\ \bibinfo {author} {\bibfnamefont {G.}~\bibnamefont {M{\"u}ller}},\ }\bibfield  {title} {\bibinfo {title} {One-atom maser},\ }\href@noop {} {\bibfield  {journal} {\bibinfo  {journal} {Physical review letters}\ }\textbf {\bibinfo {volume} {54}},\ \bibinfo {pages} {551} (\bibinfo {year} {1985})}\BibitemShut {NoStop}%
\bibitem [{\citenamefont {Scully}\ \emph {et~al.}(1989)\citenamefont {Scully}, \citenamefont {Zhu},\ and\ \citenamefont {Gavrielides}}]{scully1989degenerate}%
  \BibitemOpen
  \bibfield  {author} {\bibinfo {author} {\bibfnamefont {M.~O.}\ \bibnamefont {Scully}}, \bibinfo {author} {\bibfnamefont {S.-Y.}\ \bibnamefont {Zhu}},\ and\ \bibinfo {author} {\bibfnamefont {A.}~\bibnamefont {Gavrielides}},\ }\bibfield  {title} {\bibinfo {title} {Degenerate quantum-beat laser: Lasing without inversion and inversion without lasing},\ }\href@noop {} {\bibfield  {journal} {\bibinfo  {journal} {Physical review letters}\ }\textbf {\bibinfo {volume} {62}},\ \bibinfo {pages} {2813} (\bibinfo {year} {1989})}\BibitemShut {NoStop}%
\bibitem [{\citenamefont {Kocharovskaya}(1992)}]{kocharovskaya1992amplification}%
  \BibitemOpen
  \bibfield  {author} {\bibinfo {author} {\bibfnamefont {O.}~\bibnamefont {Kocharovskaya}},\ }\bibfield  {title} {\bibinfo {title} {Amplification and lasing without inversion},\ }\href@noop {} {\bibfield  {journal} {\bibinfo  {journal} {Physics Reports}\ }\textbf {\bibinfo {volume} {219}},\ \bibinfo {pages} {175} (\bibinfo {year} {1992})}\BibitemShut {NoStop}%
\bibitem [{sup()}]{supplementary}%
  \BibitemOpen
  \href@noop {} {}\bibinfo {note} {See Supplemental Material at [URL will be inserted by publisher] for derivations and additional details}\BibitemShut {NoStop}%
\bibitem [{\citenamefont {Maruyama}\ \emph {et~al.}(2009)\citenamefont {Maruyama}, \citenamefont {Nori},\ and\ \citenamefont {Vedral}}]{maruyama2009physics}%
  \BibitemOpen
  \bibfield  {author} {\bibinfo {author} {\bibfnamefont {K.}~\bibnamefont {Maruyama}}, \bibinfo {author} {\bibfnamefont {F.}~\bibnamefont {Nori}},\ and\ \bibinfo {author} {\bibfnamefont {V.}~\bibnamefont {Vedral}},\ }\bibfield  {title} {\bibinfo {title} {The physics of maxwell's demon and information},\ }\href@noop {} {\bibfield  {journal} {\bibinfo  {journal} {Rev. Mod. Phys.}\ }\textbf {\bibinfo {volume} {81}},\ \bibinfo {pages} {1} (\bibinfo {year} {2009})}\BibitemShut {NoStop}%
\bibitem [{\citenamefont {Leff}\ and\ \citenamefont {Rex}(2002)}]{leff2002maxwell}%
  \BibitemOpen
  \bibfield  {author} {\bibinfo {author} {\bibfnamefont {H.}~\bibnamefont {Leff}}\ and\ \bibinfo {author} {\bibfnamefont {A.~F.}\ \bibnamefont {Rex}},\ }\href@noop {} {\emph {\bibinfo {title} {Maxwell's Demon 2 Entropy, Classical and Quantum Information, Computing}}}\ (\bibinfo  {publisher} {CRC Press},\ \bibinfo {year} {2002})\BibitemShut {NoStop}%
\bibitem [{\citenamefont {Bennett}(1987)}]{bennett1987demons}%
  \BibitemOpen
  \bibfield  {author} {\bibinfo {author} {\bibfnamefont {C.~H.}\ \bibnamefont {Bennett}},\ }\bibfield  {title} {\bibinfo {title} {Demons, engines and the second law},\ }\href@noop {} {\bibfield  {journal} {\bibinfo  {journal} {Scientific American}\ }\textbf {\bibinfo {volume} {257}},\ \bibinfo {pages} {108} (\bibinfo {year} {1987})}\BibitemShut {NoStop}%
\bibitem [{\citenamefont {Lloyd}(1989)}]{lloyd1989use}%
  \BibitemOpen
  \bibfield  {author} {\bibinfo {author} {\bibfnamefont {S.}~\bibnamefont {Lloyd}},\ }\bibfield  {title} {\bibinfo {title} {Use of mutual information to decrease entropy: Implications for the second law of thermodynamics},\ }\href@noop {} {\bibfield  {journal} {\bibinfo  {journal} {Physical Review A}\ }\textbf {\bibinfo {volume} {39}},\ \bibinfo {pages} {5378} (\bibinfo {year} {1989})}\BibitemShut {NoStop}%
\bibitem [{\citenamefont {Oberst}\ \emph {et~al.}(2007)\citenamefont {Oberst}, \citenamefont {Vewinger},\ and\ \citenamefont {Lvovsky}}]{oberst2007time}%
  \BibitemOpen
  \bibfield  {author} {\bibinfo {author} {\bibfnamefont {M.}~\bibnamefont {Oberst}}, \bibinfo {author} {\bibfnamefont {F.}~\bibnamefont {Vewinger}},\ and\ \bibinfo {author} {\bibfnamefont {A.}~\bibnamefont {Lvovsky}},\ }\bibfield  {title} {\bibinfo {title} {Time-resolved probing of the ground state coherence in rubidium},\ }\href@noop {} {\bibfield  {journal} {\bibinfo  {journal} {Optics letters}\ }\textbf {\bibinfo {volume} {32}},\ \bibinfo {pages} {1755} (\bibinfo {year} {2007})}\BibitemShut {NoStop}%
\bibitem [{\citenamefont {Jiang}\ \emph {et~al.}(2016)\citenamefont {Jiang}, \citenamefont {Zhang},\ and\ \citenamefont {Wang}}]{jiang2016electromagnetically}%
  \BibitemOpen
  \bibfield  {author} {\bibinfo {author} {\bibfnamefont {X.}~\bibnamefont {Jiang}}, \bibinfo {author} {\bibfnamefont {H.}~\bibnamefont {Zhang}},\ and\ \bibinfo {author} {\bibfnamefont {Y.}~\bibnamefont {Wang}},\ }\bibfield  {title} {\bibinfo {title} {Electromagnetically induced transparency in a zeeman-sublevels $\lambda$-system of cold 87rb atoms in free space},\ }\href@noop {} {\bibfield  {journal} {\bibinfo  {journal} {Chinese Physics B}\ }\textbf {\bibinfo {volume} {25}},\ \bibinfo {pages} {034204} (\bibinfo {year} {2016})}\BibitemShut {NoStop}%
\bibitem [{\citenamefont {Lee}\ \emph {et~al.}(2014)\citenamefont {Lee}, \citenamefont {Vrijsen}, \citenamefont {Teper}, \citenamefont {Hosten},\ and\ \citenamefont {Kasevich}}]{lee2014many}%
  \BibitemOpen
  \bibfield  {author} {\bibinfo {author} {\bibfnamefont {J.}~\bibnamefont {Lee}}, \bibinfo {author} {\bibfnamefont {G.}~\bibnamefont {Vrijsen}}, \bibinfo {author} {\bibfnamefont {I.}~\bibnamefont {Teper}}, \bibinfo {author} {\bibfnamefont {O.}~\bibnamefont {Hosten}},\ and\ \bibinfo {author} {\bibfnamefont {M.~A.}\ \bibnamefont {Kasevich}},\ }\bibfield  {title} {\bibinfo {title} {Many-atom--cavity {QED} system with homogeneous atom--cavity coupling},\ }\href@noop {} {\bibfield  {journal} {\bibinfo  {journal} {Optics letters}\ }\textbf {\bibinfo {volume} {39}},\ \bibinfo {pages} {4005} (\bibinfo {year} {2014})}\BibitemShut {NoStop}%
\end{thebibliography}%

\newpage

\begin{widetext}

\section{Supplemental Information}

\section*{S1:Atom-Cavity Interaction Hamiltonians and Photon Number Dynamics}

We consider a cavity quantum electrodynamics (QED) setup where atoms interact with a single-mode photon field through different internal level configurations. In this section, we derive the effective photon-number dynamics under three atomic configurations: (1) a one-excited-state and $N$-ground-state configuration, (2) a two-excited-state and one-ground-state configuration, and (3) a two-ground-state and two-excited-state configuration.

\subsection*{Case I: One-excited-state and $N$-ground-state configuration}

The $k$th cavity photon mode, with frequency $\nu_k$, is represented by the annihilation and creation operators $\hat{a}_k$ and $\hat{a}^\dagger_k$, respectively. The ground states $|g_i\rangle$ are degenerate, and the energy gap between the excited state $|e\rangle$ and the ground states corresponds to $\hbar\omega$. In the interaction picture, applying the rotating-wave approximation, the Hamiltonian is expressed as
\begin{eqnarray}
\hat{V}(t) =  \sum_k \sum_{i=1}^N \lambda_{k}^* \hat{a}_k e^{i(\omega - \nu_k)t} \hat{\sigma}_{e g_i}+ \sum_k \sum_{i=1}^N \lambda_{k} \hat{a}^{\dag}_k e^{-i(\omega - \nu_k)t} \hat{\sigma}_{g_i e},
\label{intehami}
\end{eqnarray}
where $\sigma_{e g_i}=|e\rangle\langle g_i|$ and $\sigma_{g_i e}=|g_i\rangle\langle e|$. Here, $\lambda_k$ denotes the coupling strength between the $k$th photon mode and the atom. Assuming the thermal baths are sufficiently large, the back-action of the heating and cooling processes on them is negligible. We can therefore apply the Born approximation, which allows us to focus solely on the density operator $\rho$ of the atom-cavity system. The equation of motion for $\rho$, obtained using second-order perturbation theory (valid in the weak-coupling regime), is given by
\begin{eqnarray}
\dot{\rho}(t) = -\frac{i}{\hbar} \left[ \hat{V}(t), \rho(t_0) \right] - \frac{1}{\hbar^2} \int_{t_0}^{t} \left[ \hat{V}(t), \left[ \hat{V}(t'), \rho(t') \right] \right] dt'.
\label{dynamical}
\end{eqnarray}
Here, $t_0$ denotes the initial time at which the system starts evolving. We note that for a thermal state, the expectation values are $\langle \hat{a}_k \rangle = \langle \hat{a}_k^\dagger \rangle = 0$, $\langle \hat{a}_k \hat{a}_{k'} \rangle = \langle \hat{a}_k^\dagger \hat{a}_{k'}^\dagger \rangle = 0$, $\langle \hat{a}_k^\dagger \hat{a}_{k'} \rangle = \bar{n}_k \delta_{kk'}$, and $\langle \hat{a}_k \hat{a}_{k'}^\dagger \rangle = (\bar{n}_k + 1) \delta_{kk'}$. Therefore, the first term in Eq.~(\ref{dynamical}) is zero. Moreover, in a one-dimensional photon gas confined within a cavity, when the cavity length $L$ is sufficiently large and the thermal energy $k_\mathrm{B} T$ is much greater than the energy spacing between photon modes, the cavity spectrum can be approximated as continuous. In this case, we have $\sum_k \to \frac{L}{\pi} \int_{0}^{\infty} dk$. Substituting this into Eq.~(\ref{dynamical}), the dynamical equation for the density operator simplifies to the following form:
\begin{eqnarray}
    \dot{\rho}(t) & = & - \frac{1}{\hbar^2} \int_{t_0}^{t} dt' \left( \frac{L}{\pi} \right)^2 \int_0^{\infty} dk \, \lambda_k^2 \cr
    && \times \left\{ e^{i(\omega - \nu_k)(t - t')} \left[ \hat{a}_k \sum_{i=1}^N \hat{\sigma}_{e g_i}, \left[ \hat{a}^\dagger_{k} \sum_{j=1}^N \hat{\sigma}_{g_j e}, \rho(t') \right] \right] \right.\cr
    &&\left. + e^{-i(\omega - \nu_k)(t - t')} \left[ \hat{a}^\dagger_k \sum_{i=1}^N \hat{\sigma}_{g_i e}, \left[ \hat{a}_{k} \sum_{j=1}^N \hat{\sigma}_{e g_j}, \rho(t') \right] \right] \right\}.
\label{rhosimplify}
\end{eqnarray}
In the present work, we study the steady-state operation of the cavity-atom system. Assuming that the density matrix varies slowly with time, we invoke the Markov approximation. Moreover, under the assumption of weak-coupling strength $\lambda_k$, the product state approximation is justified. Consequently, we can factorize the total density matrix as $\rho(t') \approx \rho(t) \approx \rho_A(t) \otimes \rho_C(t)$, where $\rho_A(t)$ and $\rho_C(t)$ denote the density matrices of the atom and the cavity, respectively. By performing the time integration $\int_{t_0}^{t} dt' e^{i(\omega - \nu_k)(t - t')} = \pi \delta(\omega - \nu_k)$, and assuming that the resonance condition $\nu_{k_0} = \omega$ is satisfied for the $k_0$th mode, we infer that only the cavity mode with frequency $\nu_{k_0}=\omega$ is excited. Substituting this result into Eq.~(\ref{rhosimplify}) and tracing out the atomic degrees of freedom, we obtain the dynamical equation for the reduced cavity density operator:

\begin{eqnarray}
    \dot{\rho}_C & = & - \frac{2\lambda^2_{k_0} L^2} {\pi\hbar^2}  
    \left[ N P_{ee} \left(a_{k_0} a^\dag_{k_0}\rho_C +\rho _Ca_{k_0}a^\dag_{k_0} - 2a^\dag_{k_0}\rho_C a_{k_0} \right) \right.\cr
    && \left.+ \sum_{i,j}P_{g_i g_{j}} \left(a^\dag_{k_0} a_{k_0}\rho_C + \rho_C a^\dag_{k_0} a_{k_0} - 2a_{k_0}\rho_C a^\dag_{k_0}  \right) \right].
    \label{reduced}
\end{eqnarray}
By defining $\alpha=2\lambda^2_{k_0} L^2/ {\pi\hbar^2}$, and using expression~(\ref{reduced}), the equation for the average cavity photon number $\bar{n}_\mathrm{Q}$ resulting from quantum coherence is given by:
\begin{eqnarray}
\dot{\bar{n}}_\mathrm{Q} = \alpha \left[NP_{ee} (\bar{n}_\mathrm{Q}+1)-\sum_{i,j}P_{g_i g_{j}}\bar{n}_\mathrm{Q}\right].
\end{eqnarray}
The steady-state solution ($\dot{\bar{n}}_\mathrm{Q}=0$) yields
\begin{eqnarray}
\bar{n}_\mathrm{Q} &=& \frac{NP_{ee}}{\sum_{i,j}P_{g_i g_{j}}-N P_{ee}} \nonumber \\
&=& \frac{\bar{n}} {(1+ \bar{n}\epsilon_g + \epsilon_g)} ,
\end{eqnarray}
where the second equality follows from the assumption $P_{g_i g_i} = p$, using the definitions $\bar{n} = (p/P_{ee}-1)^{-1}$ and $\epsilon_g={\sum_{i\neq j} P_{g_i g_j}}/{\sum_{i} P_{g_i g_i}}$. Correspondingly, the effective cavity temperature, defined via the Boltzmann distribution, is
\begin{equation}
    T_\mathrm{Q}= \frac{\hbar \omega }{k_\mathrm{B} \ln[(1+\epsilon_g) (1+\bar{n}^{-1})]}.
\end{equation} 

\subsection*{Case II: Two-excited-state and one-ground-state configuration}

For a three-level atom with two excited states $|e_1\rangle$, $|e_2\rangle$ and one ground state $|g\rangle$, the interaction Hamiltonian is given by
\begin{eqnarray}
\hat{V}(t) =  \sum_k \sum_{j=1,2} \lambda_{k}^* \hat{a}_k e^{i(\omega - \nu_k)t} \hat{\sigma}_{e_j g} + \sum_k \sum_{j=1,2} \lambda_{k} \hat{a}^{\dag}_k e^{-i(\omega - \nu_k)t} \hat{\sigma}_{g e_j}.
\end{eqnarray}
The corresponding master equation becomes
\begin{eqnarray}
\dot{\rho}(t) = -\alpha \sum_{i,j=1,2} P_{e_i e_j} \left(a a^\dagger \rho + \rho a a^\dagger - 2a^\dagger \rho a \right) + 2 \alpha P_{gg} \left( a^\dagger a \rho + \rho a^\dagger a - 2a \rho a^\dagger \right).
\end{eqnarray}
The photon-number dynamics is
\begin{align}
    \dot{\bar{n}}_\mathrm{Q} = 2\alpha \left[ \sum_{i,j=1,2} P_{e_i e_j} (\bar{n}_\mathrm{Q}+1) - 2P_{gg}\,\bar{n}_\mathrm{Q} \right] ,
    \label{meaneq2}
\end{align}
and in steady state ($\dot{\bar{n}}_\mathrm{Q}=0$) one finds 
\begin{eqnarray}
\bar{n}_\mathrm{Q} =\frac{\bar{n}(1+\epsilon_e)}{1 - \bar{n}\epsilon_e}.
\end{eqnarray}
The corresponding steady state cavity temperature is
\begin{equation}
T_\mathrm{Q} = \frac{\hbar \omega}{k_\mathrm{B} \ln\left( \frac{\bar{n} + 1}{\bar{n}(1+\epsilon_e)} \right)}.
\label{temperature}
\end{equation}

\subsection*{Case III: Unified four-level configuration}

For the four-level configuration with two excited states and two ground states, the Hamiltonian takes the form
\begin{eqnarray}
\hat{V}(t) =  \sum_k \sum_{i,j=1,2} \lambda_{k}^* \hat{a}_k e^{i(\omega - \nu_k)t} \hat{\sigma}_{e_i g_j} + \sum_k \sum_{i,j=1,2} \lambda_{k} \hat{a}^{\dag}_k e^{-i(\omega - \nu_k)t} \hat{\sigma}_{g_i e_j}.
\end{eqnarray}
The resulting master equation is
\begin{eqnarray}
\dot{\rho}(t) = -\alpha \sum_{i,j=1,2} P_{e_i e_j} \left(a a^\dagger \rho + \rho a a^\dagger - 2a^\dagger \rho a \right) + \alpha  \sum_{i,j=1,2} P_{g_i g_j} \left( a^\dagger a \rho + \rho a^\dagger a - 2a \rho a^\dagger \right).
\end{eqnarray}
The photon-number dynamics becomes
\begin{equation}
\dot{\bar{n}}_\mathrm{Q} = 2\alpha \left[ \sum_{i,j=1,2} P_{e_i e_j} (\bar{n}_\mathrm{Q}+1) - \sum_{i,j=1,2}P_{g_ig_j} \bar{n}_\mathrm{Q} \right] .
\label{meaneq3}
\end{equation}
The steady-state solution yields
\begin{equation}
\bar{n}_\mathrm{Q} = \frac{\bar{n}(1+\epsilon_e)}{1 - \bar{n}\epsilon_e + (\bar{n}+1)\epsilon_g},
\label{neq4}
\end{equation}
with the corresponding effective temperature
\begin{equation}
T_\mathrm{Q} = \frac{\hbar \omega}{k_B \ln \left( \dfrac{(\bar{n}+1)(1+\epsilon_g)}{\bar{n}(1+\epsilon_e)} \right)}.
\label{tqvsepsilon4}
\end{equation}
For the case $\epsilon_g=0$, the four-level setup reduces to the two-excited-state and one-ground-state configuration. Consequently, Eq.~(\ref{tqvsepsilon4}) recovers Eq.~(\ref{temperature}), consistent with the result derived directly from photon-number dynamics.

\section*{S2. Bounds and constraints}
\label{bounds}
In this section we summarize the physical constraints on the coherence parameters that ensure the atomic density matrix remains positive semidefinite and the steady-state photon number is positive. These bounds determine the allowed ranges of $\epsilon_g$ and $\epsilon_e$ used in the main text.

\subsection*{Case I: One-excited-state and $N$-ground-state configuration}
Assume that the coherence between any pair of distinct ground states is equal to a constant value $\xi$, independent of the number of degenerate ground states, $N$. Thus, the atom’s density matrix takes the form
\begin{align}
    \rho_A
= P_{ee}|e\rangle\langle e|
+ p \sum_{i=1}^{N} |g_i\rangle\langle g_i|
+ \xi \sum_{\substack{i\neq j }}^{N} |g_i\rangle\langle g_j| .
\end{align}
The density matrix form is
\begin{eqnarray}
\rho_A = \begin{pmatrix}
P_{ee} & 0 & 0 & \cdots & 0 \\
0 & p & \xi & \cdots & \xi \\
0 & \xi & p & \cdots & \xi \\
\vdots & \vdots & \vdots & \ddots & \vdots \\
0 & \xi &\xi & \cdots & p
\end{pmatrix},
\end{eqnarray} 
where the $N\times N$ block corresponding to the ground-state manifold has identical off-diagonal entries $\xi$. This block can be written compactly as
\begin{equation}
M = (p-\xi)I_N + \xi J_N,
\end{equation}
with $I_N$ the $N\times N$ identity matrix and $J_N$ the $N\times N$ all-ones matrix.

We first consider the case where the coherence terms among the ground states take their maximum possible values. Since the atom’s reduced density matrix $M$ must be positive semidefinite, all of its principal submatrices must also be positive semidefinite. In particular, the $N\times N$ ground-state block must have a non-negative determinant,
\begin{equation}
\det[M] \ge0,
\end{equation}
which yields 
\begin{equation}
-\frac{ p}{N-1}\;\le\; \xi \;\le\; p.
\end{equation}
From the positivity of $\bar{n}_\mathrm{Q}$ in Eq.~(\ref{nqn}), we find that the tightest lower bound on $\xi$ is $-\frac{p}{(N-1)(\bar{n}_\mathrm{h}+1)}<\xi$. Therefore, the range of $\xi$ under consideration is
\begin{equation}
    -\frac{p}{(N-1)(\bar{n}_\mathrm{h}+1)}<\xi<p.
\end{equation}
Correspondingly, the range of $\chi$ and $\epsilon_g$ are
\begin{equation}
    -\frac{1}{(N-1)(\bar{n}_\mathrm{h}+1)}<\chi<1.
\end{equation}
\begin{equation}
    -\frac{1}{\bar{n}_\mathrm{h}+1}<\epsilon_g<N-1.
\end{equation}

\subsection*{Case II: Two-excited-state and one-ground-state configuration}
The excited-state coherence parameter $\epsilon_e$ is defined as
\begin{equation}
\epsilon_e = \frac{ P_{e_1 e_2}+ P_{e_2 e_1} }{P_{e_1e_1} + P_{e_2e_2}}.
\end{equation}
Since $P_{e_1e_1} = P_{e_2e_2} \ge |P_{e_1e_2}|$, it follows that $-1 \leq \epsilon_e < 1$.

Requiring the steady-state photon number,
\begin{equation}
\bar{n}_\mathrm{Q} = \frac{\bar{n}(1+\epsilon_e)}{1 - \bar{n}\epsilon_e},
\end{equation}
to remain positive yields the tighter bound
\begin{equation}
-1 \leq \epsilon_e < \frac{1}{\bar{n}}.
\end{equation}

\subsection*{Case III: Unified four-level configuration}
The atomic density matrix is
\begin{eqnarray}
\rho_A = \begin{pmatrix}
P_{e_1e_1} & P_{e_1e_2} & 0  & 0 \\
P_{e_2e_1} & P_{e_2e_2} & 0 & 0 \\
0 & 0 & P_{g_1 g_1} & P_{g_1 g_2} \\
0 & 0 & P_{g_2 g_1} & P_{g_2 g_2}
\end{pmatrix}.
\label{matrix}
\end{eqnarray}
Assuming $P_{g_1 g_1} = P_{g_2 g_2}=p$, we then have 
$P_{e_1 e_1} = P_{e_2 e_2}=1/2-p$. Further imposing 
$P_{e_1 e_2}=P_{e_2 e_1}$ and $P_{g_1 g_2}=P_{g_2 g_1}$, 
the coherence parameters are
\[
\epsilon_e=\frac{P_{e_1 e_2}}{1/2-p}, \qquad 
\epsilon_g=\frac{P_{g_1 g_2}}{p},
\]
consistent with the definitions given in the main text. 
The determinant of the density matrix is then
\begin{equation}
\det[\rho_A]=\left(\tfrac{1}{2}-p\right)^2 p^2 (1-\epsilon_e^2)(1-\epsilon_g^2),
\end{equation}
which enforces the positivity constraints $|\epsilon_e|\leq 1$ and 
$|\epsilon_g|\leq 1$, which are automatically satisfied under our assumptions.

\section*{S3. Quantum Carnot efficiency}
In this section we derive the expressions for the quantum Carnot efficiency corresponding to the heating and cooling regimes discussed in the main text. The results show how coherence modifies the effective temperatures and, consequently, the efficiency of the cycle.

For the case where ground-state coherence increases the effective temperature during the isothermal expansion process (that is, the case where \(\epsilon_g=\chi (N-1)<0\)), the quantum Carnot efficiency is 
\begin{eqnarray}
    \eta_\mathrm{Q}&=&1-\frac{T_\mathrm{c}}{T_\mathrm{Q}}\cr
    &=&\eta-T_\mathrm{c} \left(\frac{1}{T_\mathrm{Q}}-\frac{1}{T_\mathrm{h}}\right)\cr
    &=&\eta-\frac{\ln(1+\epsilon_g)}{\ln(1+\bar{n}_\mathrm{c}^{-1})}.
\end{eqnarray} 
Assuming a single-bath case with $T_\mathrm{bath}=T_\mathrm{c}=T_\mathrm{h}$, so that $\bar{n}_\mathrm{eq}=\bar{n}_\mathrm{h}=\bar{n}_\mathrm{c}$, the corresponding quantum efficiencies reduce to
\begin{equation}
    \eta_\mathrm{Q}=-\frac{\ln(1+\epsilon_g)}{\ln(1+\bar{n}_\mathrm{eq}^{-1})}=-\frac{\ln[1+\chi(N-1)]}{\ln(1+\bar{n}_\mathrm{eq}^{-1})}.
\end{equation} 
For the case where ground-state coherence decreases the effective temperature during the isothermal compression process (that is, the case where \(\epsilon_g=\chi (N-1)>0\)), the quantum efficiency is 
\begin{eqnarray}
    \eta_\mathrm{Q}&=&1-\frac{T_\mathrm{Q}}{T_\mathrm{h}}\cr
    &=&\eta+\frac{T_\mathrm{h}}{T_\mathrm{c}} \frac{\ln(1+\epsilon_g)}{\ln(1+\bar{n}^{-1}_\mathrm{Q})}\cr
    &=&\eta+\frac{T_\mathrm{h}}{T_\mathrm{c}} \frac{\ln(1+\epsilon_g)}{\ln[(1+\epsilon_g)(1+\bar{n}^{-1}_\mathrm{c})]}.
\end{eqnarray} 
with $T_\mathrm{bath}=T_\mathrm{c}=T_\mathrm{h}$ assumed,
\begin{equation}
\eta_\mathrm{Q}=\left[1+\frac{\ln(1+\bar{n}_\mathrm{eq}^{-1})}{\ln(1+\epsilon_g)}\right]^{-1}=\left[1+\frac{\ln(1+\bar{n}_\mathrm{eq}^{-1})}{\ln[1+\chi(N-1)]}\right]^{-1}.
\end{equation} 

\end{widetext}

\end{document}